\documentstyle{article}
\begin{document}
\LARGE
\begin{center}
\bf  Pair Creation of Black Holes in Anti-de Sitter Space
Background 
 
\vspace*{0.6in}
\normalsize \large \rm 
Wu Zhong Chao

Dept. of Physics

Beijing Normal University

Beijing 100875, China

(Oct.4, 1998)

\vspace*{0.4in}
\large
\bf
Abstract
\end{center}
\vspace*{.1in}
\rm
\normalsize
\vspace*{0.1in}

In the absence of a general no-boundary proposal for
open creation, the complex constrained instanton is used as the
seed for the open pair creations of black holes
in the Kerr-Newman-anti-de Sitter family. The relative creation
probability of the chargeless and nonrotating black hole pair is
the exponential of the negative of the entropy, and that of the
charged and (or) rotating black hole pair is the exponential of
the negative of one quarter of the sum of the outer and inner
black hole horizon areas.

\vspace*{0.3in}

PACS number(s): 98.80.Hw, 98.80.Bp, 04.60.Kz, 04.70.Dy


\vspace*{0.3in}

e-mail: wu@axp3g9.icra.it

\pagebreak

\vspace*{0.3in}

In the No-Boundary Universe, the wave function of a closed
universe is  defined as
a path integral over all compact 4-metrics with matter fields
[1]. The dominant contribution to
the path integral is from the stationary action solution. At
the $WKB$ level, the wave function
can be written as
\begin{equation}
\Psi \approx e^{- I},
\end{equation}
where $I= I_r + iI_i$  is the complex action of the solution.

The Euclidean action  is
\begin{equation}
I = - \frac{1}{16 \pi} \int_M (R - 2 \Lambda + L_m) - \frac{1}{8
\pi}
\oint_{\partial M } K,
\end{equation}
where $R$ is the scalar curvature of the spacetime $M$, $\Lambda$
is the cosmological constant, $K$
is the trace of the second form of the boundary $\partial M$, and
$L_m$ is the Lagrangian of the matter content.

The imaginary part $I_i$ and real part $I_r$ of the
action represent the Lorentzian
and Euclidean evolutions in real time and imaginary time,
respectively. When their orbits are intertwined, they are
mutually perpendicular in the configuration space with the
supermetric. The probability of a Lorentzian orbit remains
constant during the evolution. One can identify the probability,
not only as the probability of the universe created, but also as
the probabilities for other Lorentzian universes obtained through
an analytic continuation from it [2].

An instanton is defined as a stationary action orbit and
satisfies the Einstein equation everywhere. It is
the seed for the creation of the universe. However, very few
regular instantons exist. The
framework of the No-Boundary Universe is much wider than that of
the instanton theory. Therefore, in order not to exclude many
interesting phenomena from
the study, one has to appeal
to the concept of constrained instantons [3]. Constrained
instantons are the orbits
with an action which is stationary
under some restriction. The restriction can be imposed
on a spacelike 3-surface of the created Lorentzian
universe. The restriction is that the 3-metric and matter
content are given at the 3-surface. The relative creation
probability from the
instanton  is the exponential of the negative of the
real part of the instanton action.

The usual prescription for finding a constrained instanton is to
obtain a complex solution to the Einstein equation and other
field equations in the complex domain of spacetime coordinates.
If there is no singularity in a compact section of the solution,
then the compact section of the solution is considered as an
instanton. If there exist singularities in the section, then the
action of the section is not stationary. The action may
only be stationary with respect to the variations under some
restrictions mentioned above. If this is the case, then the
section is a
constrained gravitational instanton. To find the constrained
instanton, one has to closely investigate the singularities. The
stationary action condition is crucial to the validation of the
$WKB$ approximation. We are going to work at the $WKB$ level for
the problem of quantum creation of a black hole pair.

A main unresolved problem in quantum cosmology is to
generalize the no-boundary proposal for an open universe. While a
general prescription 
is not available, one can still use 
analytic continuation to obtain the $WKB$ approximation to the
wave function for open universes with some kind of
symmetry. The $S^4$ space model with $O(5)$
symmetry [4] and the $FLRW$ space model with $O(4)$ symmetry [2]
have been discussed.

In this paper, we try to reduce the symmetry further, and study
the problem of quantum pair creation of black holes in the
Kerr-Newman-anti-de Sitter family. Let us discuss the most
general case, i.e. the quantum creation
of the Kerr-Newman-anti-de Sitter black hole pair. The Lorentzian
metric of the black hole spacetime is [5]
\begin{equation}
ds^2 = \rho^2(\Delta^{-1}_r dr^2 + \Delta^{-1}_\theta d\theta^2)
+ \rho^{-2}
 \Xi^{-2}
\Delta_{\theta} \sin^2 \theta (adt - (r^2 + a^2) d\phi)^2 -
\rho^{-2} \Xi^{-2}\Delta_r  (dt - a \sin^2 \theta d \phi)^2,
\end{equation}
where
\begin{equation}
\rho^2 = r^2 + a^2 \cos^2 \theta,
\end{equation}
\begin{equation}
\Delta_r = (r^2 + a^2)(1 - \Lambda r^2 3^{-1}) - 2mr + Q^2 + P^2,
\end{equation}
\begin{equation}
\Delta_{\theta} = 1 + \Lambda a^2 3^{-1} \cos^2 \theta,
\end{equation}
\begin{equation}
\Xi = 1 + \Lambda a^2 3^{-1}
\end{equation}
and $m, a, Q$ and $P$ are constants, $m$ and $ma$ representing
mass and  angular momentum. $Q$ and $P$ are electric and
magnetic charges. The cosmological constant $\Lambda$ is
negative.

One can factorize $\Delta_r$ as follows
\begin{equation}
\Delta_r = -\frac{\Lambda}{3} (r - r_0)(r - r_1)(r - r_2)(r -
r_3),
\end{equation}
where at least two roots, say $r_0, r_1$, are complex conjugates,
and we assume $r_2$ and $r_3$ are real. If this is the case, then
$r_2$ and $r_3$ must be positive. One can identify them as the
outer 
black hole and inner black hole horizons, respectively. The roots
satisfy the following relations:
\begin{equation}
\sum_i r_i = 0,
\end{equation}
\begin{equation}
\sum_{i>j} r_i r_j = - \frac{3}{\Lambda} + a^2,
\end{equation}
\begin{equation}
\sum_{i>j>k} r_ir_jr_k = - \frac{6m}{\Lambda},
\end{equation}
\begin{equation}
\prod_i r_i = - \frac{3(a^2 + Q^2 + P^2)}{\Lambda}.
\end{equation}

The horizon areas are
\begin{equation}
A_i = 4\pi (r^2_i + a^2)\Xi^{-1}.
\end{equation}

The surface gravities of the horizons are
\begin{equation}
\kappa_i =\frac{\Lambda \prod_{j\; (j \neq i)}(r_i - r_j)}{6\Xi
(r^2_i + a^2)}.
\end{equation}

We shall concentrate on the neutral case with $Q
= P = 0$ first. The Newman-anti-de Sitter case with nonzero
electric or magnetic
charge will be discussed later.

The probability of the Kerr-anti-de Sitter black hole pair 
creation, at the $WKB$
level, is the
exponential of the negative of the action of its constrained
gravitational instanton. 

The constrained instanton is constructed from the complex version
of metric (3) by setting $\tau = it$. One can have two cuts at
$\tau = \pm \Delta \tau /2$ between the two complex horizons
$r_0, r_1$. Then one makes the $f_0$-fold cover around the
horizon $r = r_0$ and the  $f_1$-fold cover around the
horizon $r = r_1$. In
order to form a fairly symmetric Euclidean manifold, one can glue
these two cuts under the condition 
\begin{equation}
f_0 \beta_0 + f_1 \beta_1 = 0,
\end{equation}
where we set the imaginary time periods $\beta_0 = 2\pi
\kappa_0^{-1}$ and $\beta_1 = 2\pi \kappa_1^{-1}$. 

The Lorentzian metric for the black hole pair created is obtained
through analytic continuation of the time coordinate from an
imaginary to a real value at the equator. The equator is two
joint
sections $\tau = consts.$ passing these horizons. It divides the
instanton into two halves. We can impose the restriction that 
the 3-geometry characterized by the parameters $m, a, Q$ and $P$
is
given at the equator for the Kerr-Newman-anti-de Sitter family.
The parameter $f_0$ or $f_1$ is the only
degree of freedom left for the pasted manifold, since the field
equation holds elsewhere.
Thus, in order to check
whether we get a stationary action solution for the given
horizons, one only needs to see whether the above action is
stationary with respect to this parameter.
The equator where the quantum transition will occur has topology
$S^2 \times S^1$.

The action due to the horizons is
\begin{equation}
I_{i, horizon} = - \frac{\pi (r^2_i + a^2)(1 -
f_i)}{\Xi}.\;\; (i = 0,1)
\end{equation}
 
The action due to the volume is
\begin{equation}
I_v = - \frac{f_0\beta_0 \Lambda}{6\Xi^2} (r^3_1 - r^3_0 +
a^2(r_1 - r_0)).
\end{equation}

If one naively takes the exponential of the negative of half the
total action, then the exponential is not identified as the wave
function at the 
creation moment of the black hole pair. The
physical reason is that what one can observe is only the angular
differentiation, or the relative rotation of the two horizons.
This situation is similar to the case of a Kerr black hole pair
in the
asymptotically flat background. There one can only measure the
rotation of
the black hole horizon from the spatial infinity. To find the
wave function for the given mass and angular momentum one has to
make the Fourier transformation [3]
\begin{equation}
\Psi(a, h_{ij}) = \frac{1}{2 \pi}\int^{\infty}_{-\infty}
d\delta e^{i\delta
J \Xi^{-2}} \Psi(\delta, h_{ij}),
\end{equation}
where $\delta$ is the relative rotation angle for the half time
period
$f_0\beta_0/2$, which is canonically conjugate to the angular
momentum $J = ma$; and the factor $\Xi^{-2}$ is due to the
time rescaling.
The angle difference $\delta$ can be evaluated
\begin{equation}
\delta = \int_0^{f_0\beta_0/2} d\tau (\Omega_0 - \Omega_1),
\end{equation}
where the angular velocities at the horizons are
\begin{equation}
\Omega_i = \frac{a}{r^2_i + a^2}.
\end{equation}

The Fourier transformation is equivalent to adding an extra term
into the action for the constrained instanton, and then the total
action becomes
\begin{equation}
I = - \pi(r^2_0 + a^2)\Xi^{-1} - \pi(r^2_1 + a^2)\Xi^{-1}=
\pi \left ( -\frac{6}{\Lambda}+ (r^2_2 + a^2)\Xi^{-1} +(r^2_3 +
a^2)\Xi^{-1} \right ).
\end{equation}

In the derivation of the second equality in (21) one notices from
eqs. (9)(10) that the sum of all horizon areas is equal to $24\pi
\Lambda^{-1}$ for all members of the Kerr-Newman-(anti-)de Sitter
family. This fact seems coincidental, but it has a deep physical
significance.

It is crucial to note that the action is independent of the time
identification period
$f_0 \beta_0$ and therefore, the manifold obtained is qualified
as a constrained instanton.
Therefore, the relative probability of the Kerr black hole pair
creation is
\begin{equation}
P_k \approx \exp - (\pi(r^2_2 + a^2)\Xi^{-1} + \pi(r^2_3 +
a^2)\Xi^{-1}).
\end{equation}
It is the exponential of the negative of one quarter of the sum
of the outer and inner black hole horizon areas.

Now, let us turn to the charged black hole case. The vector
potential can be written as 
\begin{equation}
A =\frac{ Qr(dt - a\sin^2\theta d\phi) + P \cos \theta (a dt -
(r^2 + a^2) d\phi)}{\rho^2}.
\end{equation}

We shall not consider the dyonic case in the following.

One can closely follow the neutral rotating case for calculating
the action of the corresponding constrained gravitational
instanton. The only difference is to add one more term due to the
electromagnetic field to the action of volume. For the magnetic
case, it is
\begin{equation}
\frac{f_0\beta_0 P^2}{2\Xi^2} \left ( \frac{r_0}{r^2_0+ a^2} -
\frac{r_1}{r^2_1 + a^2} \right )
\end{equation}
and for the electric case, it is
\begin{equation}
-\frac{f_0\beta_0 Q^2}{2\Xi^2} \left ( \frac{r_0}{r^2_0+ a^2} -
\frac{r_1}{r^2_1 + a^2} \right ).
\end{equation}

In the magnetic case the vector potential determines the magnetic
charge, which is the integral  over the $S^2$ factor.
Putting all these contributions together one can find
\begin{equation}
I = - \pi(r^2_0 + a^2)\Xi^{-1} - \pi(r^2_1 + a^2)\Xi^{-1}
\end{equation}
and the relative probability of the pair creation of magnetically
charged black holes is written in the same form as eq. (22).

In the electric case, one can only fix the integral
\begin{equation}
\omega = \int A,
\end{equation}
where the integral is around the $S^1$ direction.
So, what one obtains in this way is $\Psi(\omega, a, h_{ij})$.
However, one can get the wave function $\Psi (Q,a, h_{ij})$ for
 a given electric charge through the Fourier
transformation 
\begin{equation}
\Psi (Q,a, h_{ij}) = \frac{1}{2\pi} \int^{\infty}_{-\infty} d
\omega e^{i\omega Q} \Psi
(\omega,a, h_{ij}).
\end{equation}

 The Fourier transformation
is equivalent to adding one more term to the action
\begin{equation} 
\frac{f_0\beta_0 Q^2}{\Xi^2}\left ( \frac{r_0}{r^2_0+ a^2} -
\frac{r_1}{r^2_1 + a^2} \right ).
\end{equation}

Then we obtain the same probability formula for the electrically
charged rotating black hole pair creation as for the
magnetic one. The duality between the magnetic and electric cases
is recovered [3][6][7].

The case of the Kerr-Newman black hole family can be thought of
as the limit of our case as we let $\Lambda$ approach  $0$ from
below.

Of course, if one lets the angular momentum be zero, then it is
reduced into the Reissner-Nordstr$\rm\ddot{o}$m-anti-de Sitter
black hole case. If one further lets the charge be zero, then it
is  reduced into the Schwarzschild-anti-de Sitter black hole
case.

For the Schwarzschild-anti-de Sitter black hole case, there are
only three horizons,
\begin{equation}
r_2 = 2 \sqrt{\frac{1}{|\Lambda|}} \mbox{sinh} \gamma,
\end{equation}
\begin{equation}
r_1 = \bar{r}_0 = \sqrt{\frac{1}{|\Lambda|}}(-\mbox{sinh} \gamma
- i \sqrt{3} \mbox{cosh} \gamma ),
\end{equation}
where we set
\begin{equation}
\gamma \equiv \frac{1}{3} \mbox{arcsinh} (3m |\Lambda |^{1/2}).
\end{equation}
The horizon $r = r_2$ is the black hole horizon. 

The surface gravity $\kappa_i$ of $r_i$ is [5]
\begin{equation}
\kappa_i = \frac{\Lambda}{6r_i}\prod_{j = 0,1,2, (j \neq i)} (r_i
- r_j).
\end{equation}

And the total action is 
\begin{equation}
I = - \pi ( r_0^2 + r_1^2) = \pi \left (-\frac{6}{\Lambda} +
r^2_2 \right ).
\end{equation}

Therefore, the relative probability of the pair creation of
Schwarzschild-anti-de Sitter black holes, at
the $WKB$ level, is the exponential of the negative of one
quarter of the black hole horizon area. This contrasts with
the case of pair creation of 
Schwarzschild-de Sitter black holes [3][8]. The relative creation
probability for Schwarzschild-de Sitter black holes is the
exponential of the total entropy of the universe. One quarter of
the
black hole horizon area is known to be the entropy in the 
Schwarzschild-anti-de Sitter universes [9].

One may wonder why we choose horizons $r_0$ and $r_1$ to
construct the instanton. One
can also consider those constructions involving other horizons as
the
instantons. However, the real part
of the action for our choice is always greater
than that of the other choices for the given configuration, and
the wave function or the
probability is determined by the classical orbit
with the greatest real part of the action [1]. When we dealt with
the Schwarzschild-de Sitter case, the choice of the instanton
constructed from the black hole and cosmological horizons had the
greatest action accidentally, but we
did not appreciate this earlier [3]. This
point is important. For example, if, instead we use $r_2$ and
$r_3$ for constructing the charged or rotating instanton, then
the creation probability of a universe without a black hole
would be smaller than that with a pair of black holes. This is
physically absurd.

In Euclidean quantum gravity the partition function
$Z$ is identified with the path integral under the constraints.
If the system involves the imposed quantities, namely electric
charge $Q$ or (and) angular momentum $J$,  then one has to
use the grand partition function $Z$
in grand canonical ensembles for the thermodynamics study [10].
At the $WKB$ level, one has
\begin{equation}
Z = \exp - I,
\end{equation} 
where $I$ is the effective action of a constrained instanton. The
effects of the electric charge and angular momentum have been
taken into account by the two Fourier transformations.
 
The entropy $S$ can be obtained
\begin{equation}
S = -\frac{\beta \partial}{\partial \beta} \ln Z + \ln Z,
\end{equation}
where $\beta$ is the time identification period.

Thus, the condition that $I$ is independent from $\beta$ implies
that the entropy is the negative of the action. 
One can use eq. (36) to derive the ``entropy'' and it is the
negative of the action. For compact regular instantons, the fact
that the entropy is the negative of the action is shown using
different arguments in [11]. For the open creation case, if one
naively interprets the horizon areas as the ``entropy'',
then the ``entropy'' is associated with these two complex
horizons.
Equivalently,  for the chargeless and nonrotating case one can
say that the action is identical to one
quarter of the black hole horizon area at $r_2$, or the black
hole entropy up to a constant $6 \pi \Lambda^{-1}$, as we learn
in the Schwarzschild
black hole case. For the charged or (and) rotating case, the
action is identical to one quarter of the sum of the outer and
inner black hole horizon areas up to the same constant.

Our treatment of quantum creation of the
Kerr-Newman-anti-de Sitter space family using the constrained
instanton can
be thought of as a prototype of quantum gravity for an open
system, without appealing to the background subtraction approach.
The beautiful aspect
of our approach is that even in the absence of a general
no-boundary proposal for open universes, we treat the creation of
the closed and the open universes in the same way.

It can be shown that  the probability of  the universe creation
without a black hole is greater than that with a pair of black
holes in the anti-de Sitter background.

\bf References:

\vspace*{0.1in}
\rm

1. J.B. Hartle and S.W. Hawking, \it Phys. Rev. \rm \bf D\rm
\underline{28}, 2960 (1983).

2. S.W. Hawking and N. Turok, \it Phys. Lett. \rm \bf B\rm
\underline{425}, 25 (1998), hep-th/9802030.

3. Z.C. Wu, \it Int. J. Mod. Phys. \rm \bf D\rm\underline{6}, 199
(1997), gr-qc/9801020.

4. Z.C. Wu,  \it Phys. Rev. \rm \bf D\rm
\underline{31}, 3079 (1985).

5. G.W. Gibbons and S.W. Hawking, \it Phys. Rev. \bf D\rm
\underline{15}, 2738 (1977).

6. R.B. Mann and S.F. Ross, \it Phys. Rev. \bf D\rm
\underline{52}, 2254
(1995).  

7. S.W. Hawking and S.F. Ross,  \it Phys. Rev. \bf D\rm 
\underline{52}, 5865 (1995).

8. R. Bousso and S.W. Hawking, hep-th/9807148.

9. S.W. Hawking and D.N. Page, \it Commun. Math. Phys. \rm
\underline{87}, 577 (1983).

10. S.W. Hawking,  in \it General Relativity: An Einstein
Centenary Survey, \rm eds. S.W. Hawking and W. Israel, (Cambridge
University Press, 1979).

11. G.W. Gibbons and S.W. Hawking, \it Commun. Math. Phys.
\rm \underline{66}, 291 (1979).

\end{document}